\documentclass[aps,prd,twocolumn]{revtex4}
\usepackage{amsmath}
\usepackage{epsfig}
\usepackage{subfigure}
\usepackage{float}
\usepackage{verbatim} 
\usepackage{pstricks}
\usepackage{color}
\usepackage{slashed}
\usepackage{multirow}
\usepackage{pstricks}
\usepackage{booktabs}
\usepackage{subfigure}
\usepackage{epstopdf}
\usepackage{cancel}

\renewcommand{\sec}[1]{{\bf\textit{#1}}.---}

\allowdisplaybreaks[3]
\def \bal#1\eal  {\begin{align} #1 \end{align}}
\newcommand{\be} {\begin{equation}}
\newcommand{\ee} {\end{equation}}
\newcommand{\nn} {\nonumber\\}

\newcommand{\ud} {\mathrm{d}}

\newcommand{\gi}{{\gamma}}

\newcommand{\epi}{\epsilon}
\begin{document}

{\footnotesize USTC-ICTS-18-13}

\title{Positivity bounds on vector boson scattering at the LHC}
\author{Cen Zhang}
\affiliation{
Institute of High Energy Physics, Chinese Academy of Sciences, Beijing 100049, China
}
\author{Shuang-Yong Zhou}
\affiliation{
Interdisciplinary Center for Theoretical Study, University of Science and Technology of China, Hefei, Anhui 230026, China
}

\begin{abstract}
Weak vector boson scattering (VBS) is a sensitive probe of new physics effects
in the electroweak symmetry breaking.  Currently, experimental results at the
LHC are interpreted in the effective field theory approach, where
possible deviations from the Standard Model in the quartic-gauge-boson couplings
are often described by 18 dimension-8
operators.  By assuming that a UV completion exists, we derive a new set of
theoretical constraints on the coefficients of these operators, i.e.~certain
combinations of coefficients must be positive. These constraints imply that the
current effective approach to VBS has a large redundancy: only about $2\%$ of the
full parameter space leads to a UV completion.  By excluding the remaining
unphysical region of the parameter space, these constraints provide guidance
for future VBS studies and measurements.
\end{abstract} 

%\pacs{14.80}

\maketitle

\sec{Introduction} After the discovery of the Higgs boson
\cite{Aad:2012tfa,Chatrchyan:2012xdj}, the focus of particle physics has turned
to the mechanism of electroweak symmetry breaking and beyond.  At the Large Hadron
Collider (LHC), vector boson scattering (VBS) is among the processes most
sensitive to the electroweak and the Higgs sectors.  In the Standard Model
(SM), Feynman amplitudes for longitudinally polarized weak bosons individually
grow with energy, but cancellations among diagrams involving quartic gauge
boson couplings (QGC), trilinear gauge boson couplings (TGC), and Higgs
exchange occur, and lead to a total amplitude that does not grow at large energies.
If modifications from physics beyond the Standard Model (BSM) exist, they are
likely to spoil these cancellations and lead to sizable cross section
increases.

VBS processes at the LHC can be embedded in partonic processes $qq\to VVqq$,
where $q$ is a light quark.  Both ATLAS and CMS experiments have extensively
studied this kind of signatures, and the effort will continue with future runs
of LHC.  Absent clear hints for BSM theories, these studies are based on a 
bottom-up effective field theory (EFT) approach---the SMEFT
\cite{Weinberg:1978kz, Buchmuller:1985jz,Leung:1984ni}. In this approach,
deviations in QGC independent of TGC are captured by 18 dimension-8
effective operators,
 assuming that TGC and Higgs couplings are constrained by other
processes with better experimental accuracies, such as diboson and Higgs on-shell
measurements. With this assumption, VBS measurements
at the LHC have been conveniently interpreted as constraints on
these operator coefficients,
 see, for example,
\cite{Sirunyan:2017ret,Sirunyan:2019ksz,Sirunyan:2017fvv,Sirunyan:2019der,CMS:2019iuv}
for some recent results, and \cite{cmstwiki} for a compilation of existing experimental
limits on QGC coefficients.
These constraints in turn can be matched to a variety of BSM theories.  (See,
for instance,
\cite{Fleper:2016frz,Brass:2018hfw,Giudice:2003tu,Giudice:1998ck,Han:1998sg,Hewett:1998sn,Liu:2016idz,Ellis:2017edi,Davila:2013wba}
and references therein.)  The recent high-luminosity and
high-energy LHC projection has shown that future sensitivity on
dimension-8 QGC operators at the LHC can go beyond the TeV scale \cite{Azzi:2019yne}.

However, in order to admit an ultra-violet (UV) completion, QGC operator coefficients
cannot take arbitrary values.
Recently, a novel approach has been developed to set theoretical bounds on the
Wilson coefficients of a generic EFT that can be UV completed.  
Going under the name of positivity bounds, this approach
only requires a minimum set of assumptions, which are nothing but the cherished
fundamental principles of quantum field theory such as unitarity, Lorentz
invariance, locality, and causality/analyticity of scattering amplitudes.
Using the dispersion relation of the amplitude and the optical theorem,
Ref.~\cite{Adams:2006sv} established a positivity bound in the forward
scattering limit of 2-to-2 scattering (see also \cite{Pham:1985cr, Pennington:1994kc, Ananthanarayan:1994hf, Comellas:1995hq, Dita:1998mh} for earlier discussions and applications in QCD). The bound can be computed completely within the low energy EFT and implies
that a certain combination of Wilson coefficients must be positive.
Moreover, thanks to the properties of the partial wave expansion, an infinite series of non-forward $t$ derivative
positivity bounds are derived ($t$ being the Mandelstam variable)
\cite{deRham:2017avq, deRham:2017zjm} (see also \cite{Nicolis:2009qm, Vecchi:2007na, Manohar:2008tc, Bellazzini:2014waa} for earlier discussions on non-forward positivity bounds and \cite{Bellazzini:2016xrt} for earlier discussions on the forward positivity bound for particles with spin). These positivity bounds have been used
to fruitfully constrain various gravity and particle physics theories (see, e.g., \cite{Bellazzini:2015cra, Cheung:2016yqr, Bellazzini:2016xrt, Cheung:2016wjt, Bellazzini:2017fep,
Bonifacio:2016wcb, deRham:2017imi, Bellazzini:2017bkb, deRham:2018qqo, Bonifacio:2018vzv, Bellazzini:2018paj, Distler:2006if, Bellazzini:2019xts}).

In this work, we apply this approach to the SMEFT formalism for VBS processes,
and derive a whole new set of theoretical constraints on the VBS operators.
While no bounds can be derived at $\mathcal{O}(\Lambda^{-2})$
\footnote{
		Under certain model-dependent assumptions, this approach can have implications
		on Higgs operators at dimension-6 \cite{Falkowski:2012vh,Low:2009di, Bellazzini:2014waa}.
},
we show that at $\mathcal{O}(\Lambda^{-4})$ certain sums of a linear combination of
the dimension-8 QGC coefficients and a quadratic form of the dimension-6
coefficients must be positive. Because the latter is always negative-definite,
a number of positivity constraints can be inferred solely on QGC operators.

These constraints have several features.  First, based only on the most
fundamental principles of quantum field theory, they are general and
model-independent.  In addition, they have strong impacts: the currently
allowed parameter space spanned by 18 dimension-8 coefficients will be
drastically reduced, by almost two orders of magnitude in
volume. Finally, they constrain the possible directions in which SM deviations could occur,
complementary to the experimental limits.
By revealing the physically viable region in the 18-dimensional QGC parameter
space, these constraints provide important guidance for future VBS studies. On
the other hand, if the experiments observed a parameter region that violates
the positivity bounds, it would be a very clear sign of violation of the
cherished fundamental principles of modern physics.

\sec{Effective operators}
Before deriving the positivity constraints, let us briefly describe the 
model-independent SMEFT approach to VBS processes. The approach is based on the
following expansion of the Lagrangian
\begin{flalign}
	\mathcal{L}_\mathrm{EFT}=\mathcal{L}_\mathrm{SM}
	+\sum_{d>4}\sum_i\frac{f_i^{(d)}}{\Lambda^{d-4}}O_i^{(d)}   ,
\end{flalign}
where $\Lambda$ is the typical scale of new physics. $f^{(d)}$ are the
dimensionless coefficients of the dimension-$d$ effective operators.  If the
underlying theory is known and weakly coupled, they can be determined by a
matching calculation.  It can be shown that only even-dimensional
operators conserve both baryon and lepton numbers \cite{Degrande:2012wf},
so we focus on dimension-6 and dimension-8 operators.
VBS processes can be affected by TGC, Higgs couplings, and QGC couplings.  At
dimension-6, modifications to TGC and $HVV$ Higgs couplings could arise
\cite{Rauch:2016pai}.  QGC couplings are also generated, but they are fully
determined by dimension-6 TGC couplings.  At dimension-8, QGC couplings could
arise independent of TGC couplings.  They are parameterized by 18 dimension-8
operators \cite{Eboli:2006wa,Degrande:2013rea,Eboli:2016kko}.

The unique feature of VBS processes is that they are sensitive to BSM effects
that manifest as anomalous QGC couplings.  One may worry that this sensitivity
is masked by possible contamination from anomalous TGC and/or Higgs couplings.
However, if these couplings are present, we may expect to first probe them
elsewhere, e.g.~in diboson production, vector boson fusion, or Higgs
production and decay measurements, which are in general measured
with a much better accuracy than VBS.  For example, TGC couplings are
constrained by $WW$ production at LEP2, which has been measured at the percent
level accuracy \cite{Schael:2013ita}.  These constraints are further improved
by the LHC data
\cite{Butter:2016cvz,Biekotter:2018rhp,Ellis:2014jta,Ellis:2018gqa,Grojean:2018dqj}.
Higgs coupling measurements at the LHC have
reached $\mathcal{O}(10\%)$ level precision, and will continue to improve with
the high-luminosity upgrade \cite{Azzi:2019yne}.  While there might still be non-negligible
dimension-6 effects in VBS \cite{Gomez-Ambrosio:2018pnl}, the projected
sensitivity to dimension-8 QGC couplings at high-luminosity and high-energy
LHC beyond the TeV scale \cite{Azzi:2019yne} suggests that VBS will continue to
be an important channel to look for possible BSM deviations in QGC.  While
a global SMEFT approach including both dimension-6 and dimension-8 operators
would be the most reliable, given the current accuracy level one could assume
that dimension-6 effects are well-constrained by other channels, and focus on
dimension-8 QGC couplings as a first step.  This is the assumptions that is
used by the experimental collaborations to set limits on QGC couplings.  Given
that the goal of this work is to provide theory guidance to experimental
analysis, we will adopt the same assumption.  We nevertheless emphasize that our
main conclusion, i.e.~positivity bounds on QGC couplings, are independent of
the presence of dimension-6 operators, as we will explain later.  Therefore
they will continue to be useful for future global fits including dimension-6 effects.

Dimension-8 QGC couplings are described by 18 operators.
Conventionally, they are divided into three categories: S-type operators
involve only covariant derivatives of the Higgs, M-type operators include a mix
of field strengths and covariant derivatives of the Higgs, and T-type operators
include only field strengths. 
We use the convention of \cite{Degrande:2013rea} that has become standard in this
community.  The definition of these operators can be found in Eqs.~(13)-(31) of
\cite{Degrande:2013rea} ($O_{M,6}$ is redundant \cite{Sekulla}),
and we also list them in the Appendix. The 18 operator coefficients are denoted as
\begin{flalign}
	&f_{S,0},\ f_{S,1},\ f_{S,2},\ f_{M,0},\ f_{M,1},\ f_{M,2},\ f_{M,3},\ 
f_{M,4},\ f_{M,5},
	\nonumber\\
	&f_{M,7},\ f_{T,0},\ f_{T,1},\ f_{T,2},\ f_{T,5},\ f_{T,6},\ f_{T,7},\ f_{T,8},\ 
	f_{T,9}.
	\nonumber
\end{flalign}
A summary of existing experimental constraints on these coefficients can be
found in \cite{cmstwiki}.
See also Ref.~\cite{Green:2016trm} for a review of QGC measurements at the LHC
and their interpretation in the SMEFT.

\sec{Positivity bounds}  
The simplest positivity bound can be obtained by considering an elastic
scattering amplitude in the forward limit $A(s)=A(s,t=0)$~\cite{Adams:2006sv}.
Thanks to the dispersion relations, optical theorem and Froissart bound
\cite{Froissart:1961ux},
it can be shown that the second derivative of $A(s)$
w.r.t.~$s$ is positive, after subtracting contributions from the low energy
poles. In the following we shall briefly review the forward limit positivity bound,
adapted to the context of VBS.
We assume that the contributions from the higher dimensional operators are well approximated by the tree level, which is a reasonable assumption given that perturbativity in EFT is always needed for a valid analysis.

If the UV completion is weakly coupled,  
the BSM amplitude is usually well approximated by its leading tree level
contribution $A_{\rm tr}$, which is analytic and satisfies the Froissart bound.
Its BSM part simply comes from one particle exchange between SM currents. We
can derive a dispersion relation for $A_{\rm tr}$: 
\bal
&~~~~f_{\rm tr}(s_p) \equiv\frac{1}{2\pi i} \oint_{\cal C} \ud s \frac{A_{\rm tr}(s)}{(s-s_p)^3}
\\
\label{eq:lhcAtr}
&= \int^\infty_{\Lambda_{\rm th}^2\!+M^2}\!\! \frac{\ud s}{\pi}  \frac{{\rm Im} A_{\rm tr}(s) }{( s+s_p\!-\!M^2 )^3}  + \int^\infty_{\Lambda_{\rm th}^2}\!\! \frac{\ud s}{\pi}   \frac{ {\rm Im} A_{\rm tr}(s)  }{(s-s_p)^3}  ,
\eal
where $M^2\equiv 2m_1^2+2m_2^2$, $m_1$ and $m_2$ being the masses of the
interacting particles, and $\Lambda_{\rm th}(\gg M)$ is the mass of the
lightest heavy state. ${\cal C}$ is a contour that encircles
all the poles in the low energy EFT and, by analyticity of the $s$ complex
plane, can be deformed to run around the $s>\Lambda_{\rm th}^2$ and
$s<-\Lambda_{\rm th}^2$ parts of the real axis and along the infinite
semi-circles; the infinite semi-circle contributions vanish due to the
Froissart bound, and the discontinuities along the real axis give rise to ${\rm
Im} A_{\rm tr}(s,0)$ which is nonzero due to the heavy particle poles. 
Also we have restricted to crossing symmetric
amplitudes for simplicity, and to obtain the first term of
Eq.~(\ref{eq:lhcAtr}) we have made a variable change $s\to M^2-s$ and used the
crossing symmetry ${\rm Im} A_{\rm tr}(M^2 - s)={\rm Im} A_{\rm tr}(s)$.  
By the cutting rules, ${\rm Im} A_{\rm tr}(s)$ can be written as a sum of
complete squares of 2-to-1 amplitudes, and thus ${\rm Im} A_{\rm
tr}(s)>0$. Therefore we infer that $f_{\rm tr}(s_p)>0$
for $-\Lambda_{\rm th}^2<s_p<\Lambda_{\rm th}^2$. 
Due to analyticity of the amplitude in complex $s$ plane, $f_{\rm tr}(s_p)$ can
be calculated within the SMEFT as the second derivative of the effective
amplitude $A_{\rm tr}(s)$ with the poles subtracted. 
Since the SM at tree level makes no contribution to the r.h.s.~of
	Eq.~(\ref{eq:lhcAtr}), $f_{\rm tr}(s)>0$ directly gives positivity
	constraints on the Wilson coefficients.

The above argument can be easily generalized to cases where the leading EFT
amplitude is matched to the loop amplitude in the full theory, and one can derive
positivity for the lowest order $n$-loop BSM contribution to VBS, with the SM
contribution removed.  In this case, the discontinuity above $\Lambda_{\rm th}$ must come
from unitarity cuts that only cut the BSM particles, otherwise this amplitude
would match to both tree and loop diagrams in EFT with similar sizes,
violating the perturbativity assumption.
Using the cutting rules, the discontinuity can be written as a sum of complete
squares, thus proving positivity.

For a generic UV completion,
 consider the full amplitude including the SM contribution. The latter could give
a constant contribution to the dispersion relation at one loop. To minimize its
impact, we subtract out the branch cuts within $|s|<(\epi\Lambda)^2$ ($\epi\lesssim 1$), where the
dominant SM contribution resides.  This is done by following the improved
positivity \cite{deRham:2017imi,Bellazzini:2016xrt,deRham:2017xox} and defining:
\bal
B_{\epi\Lambda}(s_p) = A(s_p) - \int^{+(\epi\Lambda)^2}_{-(\epi\Lambda)^2}\frac{\ud s}{2\pi i}\frac{{\rm Disc}A(s)}{s-s_p}   ,
\label{eq:4}
\eal
with $M_\pm<\epsilon\Lambda <\Lambda$,
$M_\pm\equiv m_1\pm m_2$.
This subtracted amplitude has the same discontinuity as $A(s)$ above $(\epi\Lambda)^2$
and also satisfies the Froissart bound\,\footnote{Strictly speaking, the Froissart bound applies to scatterings where the external states are massive. Here we assume that this bound also applies to cases where external states are all massless such as the photon scattering. One may take the view that the photon has a very small mass. Since the extra degree of freedom associated with the mass can smoothly decouple as we take the massless limit, this procedure does not actually affect the $\gamma\gamma$ positivity bounds.}.  It is free of branch cuts for
$|s|<(\epi\Lambda)^2$, and thus one can analogously obtain a dispersion relation:
\bal
& f_{\epi\Lambda}(s_p) \equiv \frac{\ud^2 B_{\epi\Lambda}(s_p)}{2\;\ud s^2}
=\left[  \int^{-(\epi\Lambda)^2}_{-\infty} \!\!\! +\! \int_{(\epi\Lambda)^2}^{\infty} \right]
\frac{\ud s}{2\pi i} \frac{{\rm Disc}A(s)}{(s-s_p)^3}
\nn
&= \int^\infty_{\small (\epi\Lambda)^2 + M^2}\!\!\! \frac{\ud s}{\pi} \frac{ {\rm Im}A(s)}{( s+s_p \! -\!  M^2 )^3} 
+ \! \int^\infty_{\small (\epi\Lambda)^2}\!\! 
\frac{\ud s}{\pi}  \frac{ {\rm Im} A(s)}{(s-s_p)^3}  .
\label{eq:5}
\eal
Making use of the optical theorem, ${\rm Im}
A(s)=[(s-M_-^2)(s-M_+^2)]^{1/2}\sigma_t>0$ for $s> M_+^2$, where $\sigma_t$ is
the total cross section. So we have
$f_{\epi\Lambda}(s_p)>0$
for $-(\epi\Lambda)^2<s_p<(\epi\Lambda)^2$.  
Again, by contour deformation,
$f_{\epi\Lambda}(s_p)$ can be evaluated within the EFT with the
subtraction term in Eq.~(\ref{eq:4}) taken into account.
This term does not contain any tree level contribution from the higher dimensional operators, but it removes
the dominant impact from the SM loop contribution.
The remaining contribution from the SM is then suppressed by $(\epi\Lambda)^{-2}$,
and can be computed explicitly.  The reason behind is that the SM contribution
mostly comes from the
discontinuity below $\epi\Lambda$, while the BSM contribution is from
above this scale, so one can choose a $\epi\Lambda$ to subtract the dominant
SM contribution without losing positivity.  In the Appendix we
compute the remaining SM contribution in the $\gamma\gamma$ channel and show that
it is negligible even comparing with the best experimental sensitivity
currently available.

\sec{Applications}
Let us first focus on dimension-8 operators.  Applying this approach to the
scattering amplitudes of VBS in the forward limit yields a set of positivity
constraints on QGC coefficients.  As an example, we present here the constraint
from $ZZ\to ZZ$ scattering:
\begin{flalign}
&8 a_3^2 b_3^2 t_W^4
   \left(F_{S,0}+F_{S,1}+F_{S,2}\right)
   +\left[a_3^2 \left(b_1^2+b_2^2\right)\right.
\nonumber\\&
\left.+\left(a_1^2 +a_2^2\right) b_3^2\right] t_W^2
   \left(-t_W^4 F_{M,3}
   +t_W^2 F_{M,5}
   -2 F_{M,1}+F_{M,7}\right)
  \nonumber\\ &
+\left[
	\left(a_1 b_1+a_2 b_2\right){}^2+\left(a_1^2+a_2^2\right) \left(b_1^2+b_2^2\right)
\right]\left(2 t_W^8 F_{T,9}
   \right.\nonumber\\&\left.
+4 t_W^4 F_{T,7}
+8 F_{T,2}\right) +8 \left(a_1 b_1+a_2 b_2\right){}^2
   \left[t_W^4 \left(t_W^4 F_{T,8}
   \right.\right.\nonumber\\&\left.\left.
	   +2 F_{T,5}+2 F_{T,6}\right)
   +4 F_{T,0}+4
   F_{T,1}\right]\ge0    ,
   \label{eq:zz}
\end{flalign}
where $t_W\equiv \tan\theta_W$ is the tangent of the weak angle. We have rewritten
the coefficients as 
\be
\label{Ftof}
F_{S,i}\equiv f_{S,i},~~
F_{M,i}\equiv e^2f_{M,i} ,~~
F_{T,i}\equiv e^4f_{T,i} .
\ee
$a_i$ and $b_i$ parametrize the polarization vectors of the two $Z$ bosons
respectively:
\begin{flalign}
	\epsilon_1^\mu=\left(a_3{p_1}/{m_Z},a_1,a_2,a_3{E_1}/{m_Z}\right)
	\\
	\epsilon_2^\mu=\left(b_3{p_2}/{m_Z},b_1,b_2,b_3{E_2}/{m_Z}\right)  ,
\end{flalign}
where real polarizations are used for simplicity.
Eq.~(\ref{eq:zz}) must hold for all real values of $a_i$ and $b_i$.
Other VBS processes yield similar but independent constraints. The full set
of results are given in the Appendix.

Interestingly, including dimension-6 operators does not change our conclusion.
If one follows the same approach and considers dimension-6 contributions, it
turns out that nontrivial constraints on them can be obtained only at the
$(f^{(6)}/\Lambda^2)^2$ level, i.e.~from diagrams involving two insertions
of operators. They always take the following form:
\begin{flalign}
	\sum_i (-x_i) \left(\sum_j y_j f_j^{(6)}\right)^2\geq0, \quad x_i>0   ,
\end{flalign}
i.e.~the sum of a set of complete square terms need to be negative.
We have checked this for all relevant dimension-6 operators in the Warsaw basis
\cite{Grzadkowski:2010es}. Explicit results are given in the Appendix.
Of course, these conditions cannot be
satisfied with dimension-6 operators alone.  Instead, it tells us that at
$\mathcal{O}(\Lambda^{-4})$ the dimension-8 contribution has to come in, with a
positive value large enough to flip the sign of the dimension-6 contribution.
Therefore, the presence of dimension-6 contributions will only make the
dimension-8 positivity constraints stronger.
Our conclusion thus holds regardless of the presence of dimension-6
operators.

It is worth mentioning that these constraints are different from bounds derived
from partial-wave unitarity \cite{Corbett:2014ora,Corbett:2017qgl}, in that
they require unitarity of the UV theory, not the low energy effective theory,
and additionally require other fundamental principles such as analyticity of
the amplitude. 
In VBS, partial-wave unitarity leads to bounds on the sizes of
$f^{(6)}/\Lambda^2$ or $f^{(8)}/\Lambda^4$, while the positivity bounds are on
the dimensionless coefficients, and lead to constraints on possible directions
of SM deviations.  These constraints are always complementary to the unitarity bounds
and experimental limits.

\sec{Physics implication}
We now describe the physics implications of our positivity constraints on
VBS processes.

\begin{table}
	\begin{tabular}{ccccccccc}
		\hline
	  $f_{S,0}$ 
	& $f_{S,1}$
	& $f_{S,2}$
	& $f_{M,0}$
	& $f_{M,1}$
	& $f_{M,2}$
	& $f_{M,3}$
	& $f_{M,4}$
	& $f_{M,5}$
		\\\hline
		+ & + & + & X & $-$ & O & $-$ &O & X
		\\\hline\hline
	  $f_{M,7}$ 
	& $f_{T,0}$
	& $f_{T,1}$
	& $f_{T,2}$
	& $f_{T,5}$
	& $f_{T,6}$
	& $f_{T,7}$
	& $f_{T,8}$
	& $f_{T,9}$
		\\\hline
		+ & + & + & + & X & + & X &+ & +
		\\\hline
	\end{tabular}
	\caption{\label{tab:one}
Positivity constraints on individual VBS operator coefficients.
$+/-$ means the coefficient must be non-negative or non-positive.
X means only $f=0$ is allowed, and O means no constraints. 
%More constraining results could be obtained by using complex polarization vectors, see Ref.~\cite{Bi:2019phv}.
}
\end{table}

First, let us turn on one operator at a time. Most experimental results
are presented as limits on individual operators, assuming all others vanish.
As shown in \cite{cmstwiki}, these limits are symmetric or nearly symmetric.
In Table~\ref{tab:one} we list our positivity constraints on individual operators.
We can see that, while $f_{M,2}$ and $f_{M,4}$ are free of such constraints,
all other coefficients are bounded at least from one side. This implies that
half of the experimentally allowed regions do not lead to a UV completion.  In
addition, $f_{M,0}$, $f_{M,5}$, $f_{T,5}$ and $f_{T,7}$ cannot individually
take any nonzero values.  $f_{M,0}$ is forbidden because the same-sign and
opposite-sign $WW$ scattering amplitudes give inconsistent constraints, while
$f_{M,5}$ is forbidden because $WW$ and $WZ$ scattering amplitudes give
inconsistent constraints. Similar situations occur for $f_{T,5}$ and $f_{T,7}$.
This implies that no UV theory could generate any of the four coefficients alone.
We will show that these conditions can be
relaxed once other coefficients are allowed to take nonzero values.  However,
the one-operator-at-a-time scenario already illustrates that the positivity
constraints have drastic impacts on the presentation and interpretation of
experimental results.

\begin{figure}[h]
	\begin{center}
		\includegraphics[width=.68\linewidth]{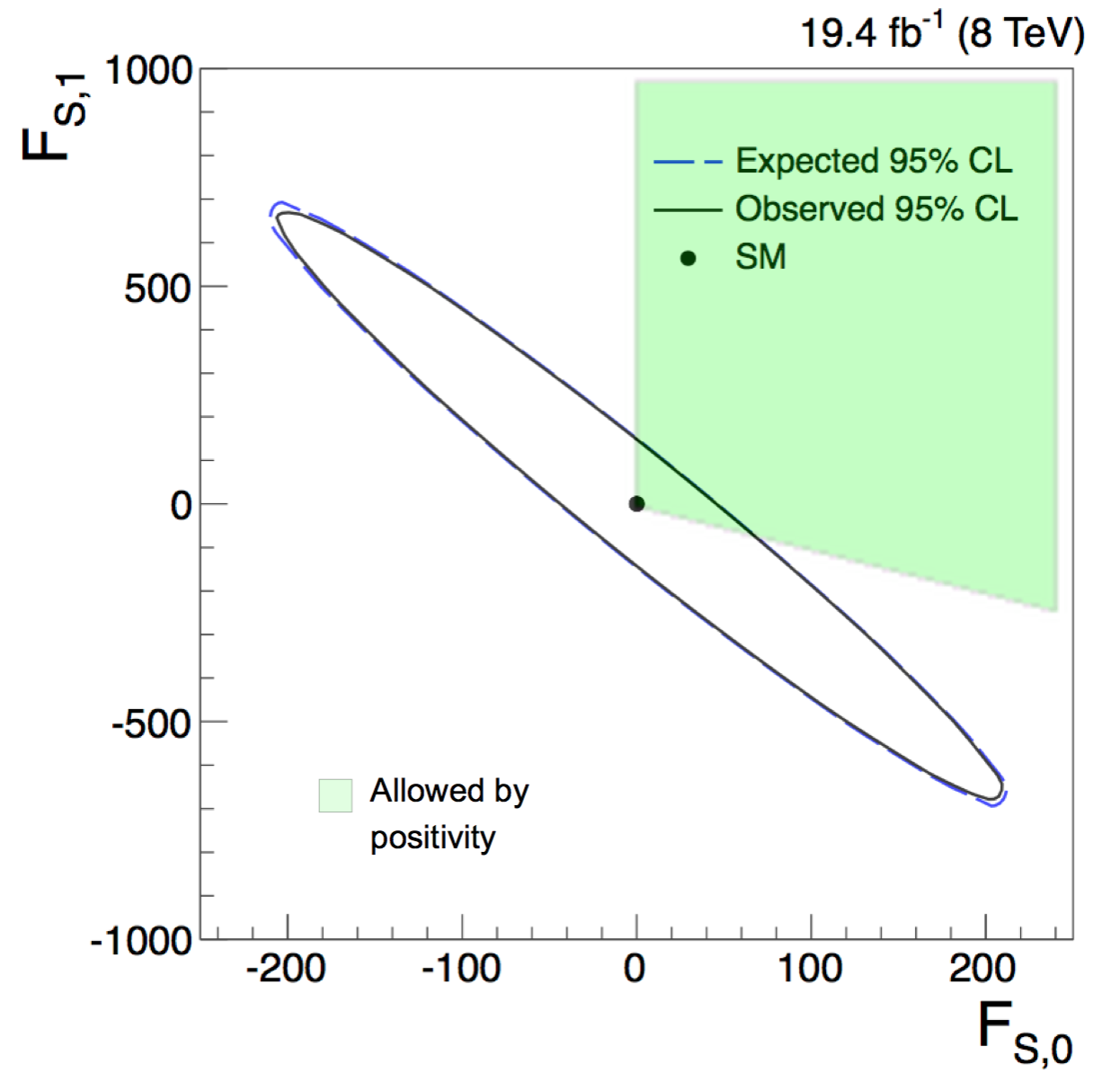}
	\end{center}
	\caption{\label{fig:cms}
	Positivity constraints on $F_{S,0}\equiv f_{S,0}$ and $F_{S,1}\equiv f_{S,0}$,
	compared with the CMS results \cite{Khachatryan:2014sta}. The green shaded
	area is allowed by positivity.
	A specific combination of $F_{S,0}$ and $F_{S,1}$ roughly
	rescales the Standard Model distribution, and so
	the measurement is insensitive to this direction.
}
	\label{fig:CMS}
\end{figure}
\begin{figure}[h]
	\begin{center}
		\includegraphics[width=.68\linewidth]{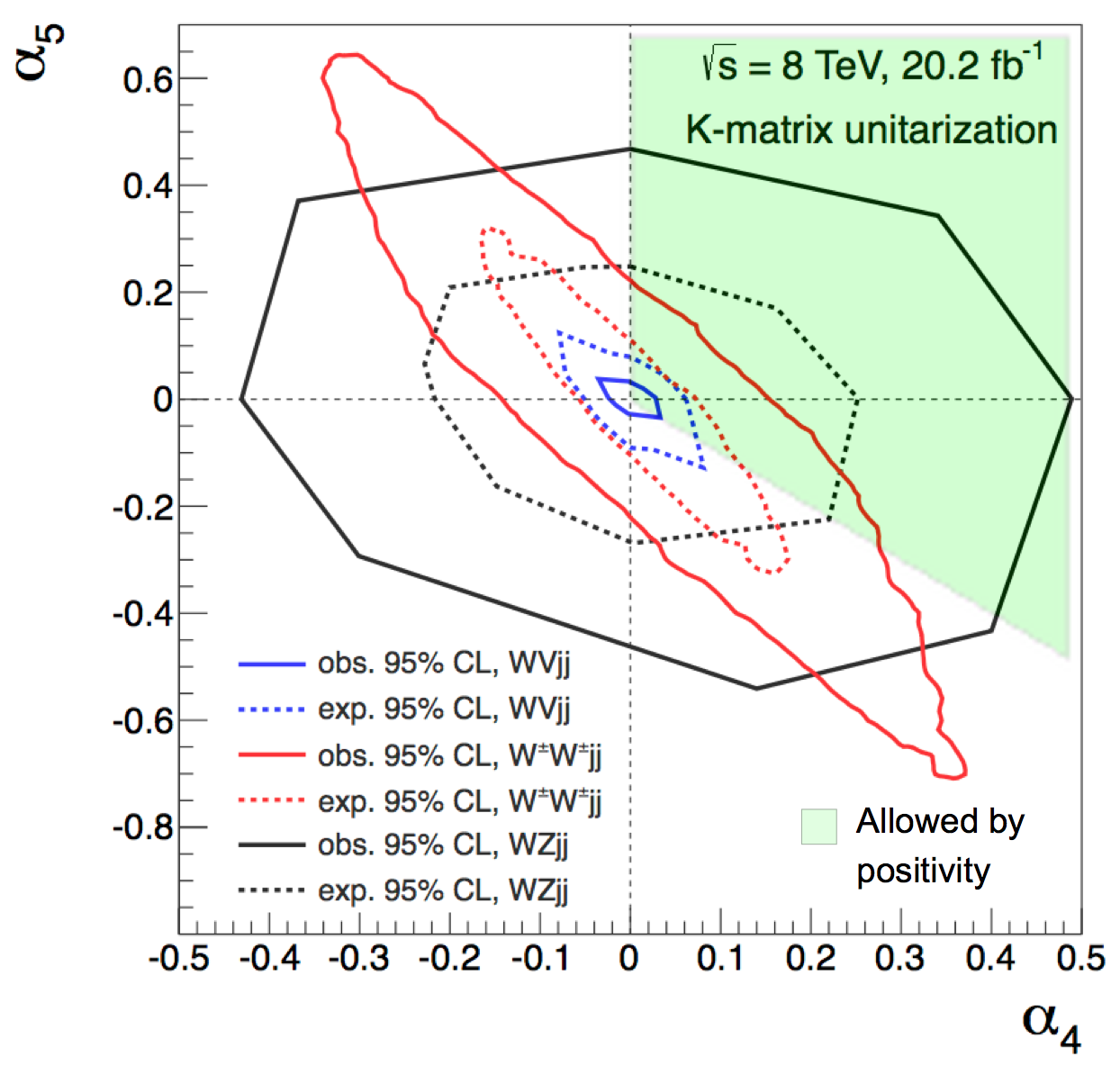}
	\end{center}
	\caption{\label{fig:atlas}
	Positivity constraints on $\alpha_4$ and $\alpha_5$, compared with the
	ATLAS results \cite{Aaboud:2016uuk,Aaboud:2016ffv,Aad:2016ett}. The green shaded area is allowed by
	positivity. The difference between the expected limits and the
	observed ones are due to fluctuations in the observed events.}
	\label{fig:CMS}
\end{figure}

Now let us turn on two operators simultaneously.  Two-operator
constraints have been presented by CMS,
on coefficients $F_{S,0}\equiv
f_{S,0}$ and $F_{S,1}\equiv f_{S,0}$ (see Eq.~(\ref{Ftof}) for
relations  between the $F$ and $f$ coefficients), and by ATLAS on
$\alpha_4$ and $\alpha_5$. The latter parameters are defined in the
nonlinear formulation, but the conversion to the linear case is
straightforward \cite{Rauch:2016pai}:

\begin{flalign}
	&\alpha_4=\frac{v^4}{16}\frac{f_{S,0}+f_{S,2}}{\Lambda^4},\quad
	\alpha_5=\frac{v^4}{16}\frac{f_{S,1}}{\Lambda^4},
\\
&\mbox{with}\quad f_{S,0}=f_{S,2}  .
\end{flalign}

In Figures~\ref{fig:cms} and
\ref{fig:atlas}, we overlay our corresponding positivity constraints on
top of the two-dimensional contour plots obtained by both experimental groups.
We can see that most of the currently allowed areas are excluded.  In other
words, only a very small fraction of the allowed parameter space could lead to
a UV completion.

\begin{figure}[h]
	\begin{center}
		\includegraphics[width=.8\linewidth]{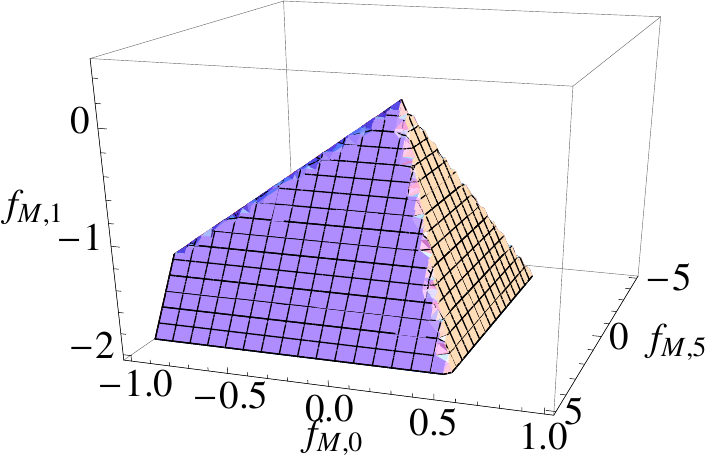}
	\end{center}
	\caption{\label{fig:3d}
Positivity constraints on $f_{M,0}$, $f_{M,1}$, and $f_{M,5}$.}
	\label{fig:CMS}
\end{figure}

We are not aware of any constraints assuming three operators are present
simultaneously. Nevertheless, for illustration, in Figure~\ref{fig:3d} we
present our constraints on three coefficients, $f_{M,0}$, $f_{M,1}$, and
$f_{M,5}$.  We can see that the allowed region has the shape of a pyramid.
Manifestly, $f_{M,0}$ and $f_{M,5}$ cannot take nonzero values alone, but this
is relaxed once $f_{M,1}$ takes a negative value.  This is consistent with our
previous observation.

Finally, a model-independent SMEFT should always take into account all operators.
An interesting question to ask in this case is the following.
Suppose future experiments at the LHC and even future colliders will collect
sufficient data to derive the global constraints on 18 operators.  How large is
the impact of the positivity constraints?

To simplify the problem, assume that all 18 operators are constrained in
the interval $-\delta<f_{i}<\delta$ without any correlations. The allowed
region in the 18-dimensional parameter space will be approximately a 18-ball
with radius $\delta$.  The fraction of its volume that satisfies all positivity
constraints is independent of $\delta$, because all the positivity constraints
are linear inequalities (see Eqs.~(\ref{boundZZ})-(\ref{boundgaga})). As we
showed previously, once the dim-6 contributions are included, they merely
provide stronger bounds, as those additional terms are all negative on the left
hand side of the inequalities.

Using a Monte Carlo integration, we
find that this fraction is $\sim2.3\%$. 
Using more generic complex polarization vectors, this fraction can be further reduced to $\sim 2.1\%$ \cite{Bi:2019phv}.  
In practice, this specific number will depend on the
relative precision achieved on each operator, but we do not expect changes of
order of magnitude.  Therefore we conclude that our positivity constraints
reduce the allowed parameter space by almost two orders of
magnitude.  

\sec{Summary} VBS processes at the LHC and future colliders are among the most
important measurements that probe the mechanism of electroweak symmetry
breaking.  We have derived a new set of constraints on the 18 QGC coefficients
in the SMEFT approach to VBS processes, by requiring that the EFT has a UV
completion. These constraints show that the current SMEFT formalism for the VBS
processes have a huge redundancy: $\sim98\%$ of the entire parameter space
spanned by 18 coefficients are unphysical and do not lead to a UV completion.

This observation provides guidance to future VBS studies.  Theoretical studies,
in particular those which employ a bottom-up approach, are advised to keep the
positivity constraints satisfied and avoid choosing unphysical benchmark
parameters.  Experimental strategies can be further optimized towards the
remaining $\sim2\%$ of the QGC parameter space.  According to the positivity
constraints, most existing limits that are symmetric can really be presented as
one-sided limits; also, individual limits on $f_{M,0}$, $f_{M,5}$, $f_{T,5}$
and $f_{T,7}$ do not have a clear physical meaning.  It is worthwhile for
future VBS measurements to take into account the positivity constraints, as
they significantly modify the prior probability densities of the QGC
coefficients by excluding unphysical values, and therefore could also affect
the resulting limits.

\section{Acknowledgements}
We would like to thank Minxin Huang, Ken Mimasu, Andrew Tolley, Eleni Vryonidou and
 Zhi-Guang Xiao for helpful discussions.  CZ is supported by IHEP under
Contract No.~Y7515540U1.
SYZ acknowledges support from the starting grant from University of Science and
Technology of China under grant No.~KY2030000089 and is also supported by National Natural Science Foundation of China under grant No.~GG2030040375.\\

\begin{widetext}

\section*{APPENDIX} 

\subsection{QGC operators and positivity bounds}

The 18 dimension-8 QGC operators discussed in this work are defined as follows:
\newcommand{\dd}[1]{D_{#1}\Phi}
\newcommand{\du}[1]{D^{#1}\Phi}
\newcommand{\dg}[1]{(#1)^\dagger}
\newcommand{\wh}{\hat{W}}
\newcommand{\bh}{\hat{B}}
\newcommand{\tr}[1]{\mathrm{Tr}\left[{#1}\right]}
\begin{flalign}
	\begin{aligned}
	\begin{array}{l}
O_{S,0}=[\dg{\dd{\mu}}\dd{\nu}]\times[\dg{\du{\mu}}\du{\nu}]
\\
O_{S,1}=[\dg{\dd{\mu}}\du{\mu}]\times[\dg{\dd{\nu}}\du{\nu}]
\\
O_{S,2}=[\dg{\dd{\mu}}\dd{\nu}]\times[\dg{\du{\nu}}\du{\mu}]
\\
O_{M,0}=\tr{\wh_{\mu\nu}\wh^{\mu\nu}}\times\left[\dg{\dd{\beta}}\du{\beta}\right]
\\
O_{M,1}=\tr{\wh_{\mu\nu}\wh^{\nu\beta}}\times\left[\dg{\dd{\beta}}\du{\mu}\right]
\\
O_{M,2}=\left[\bh_{\mu\nu}\bh^{\mu\nu}\right]\times\left[\dg{\dd{\beta}}\du{\beta}\right]
\\
O_{M,3}=\left[\bh_{\mu\nu}\bh^{\nu\beta}\right]\times\left[\dg{\dd{\beta}}\du{\mu}\right]
\\
O_{M,4}=\left[\dg{\dd{\mu}}\wh_{\beta\nu}\du{\mu}\right]\times\bh^{\beta\nu}
\\
O_{M,5}=\left[\dg{\dd{\mu}}\wh_{\beta\nu}\du{\nu}\right]\times\bh^{\beta\mu}
\\
O_{M,7}=\left[\dg{\dd{\mu}}\wh_{\beta\nu}\wh^{\beta\mu}\du{\nu}\right]
	\end{array}
\end{aligned}\quad
\begin{aligned}
	\begin{array}{l}
O_{T,0}=\tr{\wh_{\mu\nu}\wh^{\mu\nu}}\times\tr{\wh_{\alpha\beta}\wh^{\alpha\beta}}
\\
O_{T,1}=\tr{\wh_{\alpha\nu}\wh^{\mu\beta}}\times\tr{\wh_{\mu\beta}\wh^{\alpha\nu}}
\\
O_{T,2}=\tr{\wh_{\alpha\mu}\wh^{\mu\beta}}\times\tr{\wh_{\beta\nu}\wh^{\nu\alpha}}
\\
O_{T,5}=\tr{\wh_{\mu\nu}\wh^{\mu\nu}}\times\bh_{\alpha\beta}\bh^{\alpha\beta}
\\
O_{T,6}=\tr{\wh_{\alpha\nu}\wh^{\mu\beta}}\times\bh_{\mu\beta}\bh^{\alpha\nu}
\\
O_{T,7}=\tr{\wh_{\alpha\mu}\wh^{\mu\beta}}\times\bh_{\beta\nu}\bh^{\nu\alpha}
\\
O_{T,8}=\bh_{\mu\nu}\bh^{\mu\nu}\times\bh_{\alpha\beta}\bh^{\alpha\beta}
\\
O_{T,9}=\bh_{\alpha\mu}\bh^{\mu\beta}\times\bh_{\beta\nu}\bh^{\nu\alpha}   ,
	\end{array}
\end{aligned}
\end{flalign}
where
\begin{flalign}
	\wh^{\mu\nu}\equiv ig\frac{\sigma^I}{2}W^{I,\mu\nu}\,,\qquad
	\bh^{\mu\nu}\equiv ig'\frac{1}{2}B^{\mu\nu}\,.
\end{flalign}
The Lagrangian is
\begin{flalign}
	\mathcal{L}_{\rm EFT}=\mathcal{L}_\mathrm{SM}
	+\sum\frac{f_iO_i}{\Lambda^4}
\end{flalign}
and we redefine the coefficients:
\begin{flalign}
F_{S,i}\equiv f_{S,i}, \quad
F_{M,i}\equiv e^2f_{M,i},\quad
F_{T,i}\equiv e^4f_{T,i}.
\end{flalign}

The positivity constraints are derived from the crossing symmetric, forward scattering amplitude
$V_1V_2\to V_1V_2$, where $V_i=Z,W^\pm,\gamma$,
with real polarization vectors:
\begin{flalign}
	\epsilon^\mu(V_1)=\left(a_3\frac{p_1}{m_{V_1}},a_1,a_2,a_3\frac{E_1}{m_{V_1}}\right)    ,
	\\
	\epsilon^\mu(V_2)=\left(b_3\frac{p_2}{m_{V_2}},b_1,b_2,b_3\frac{E_2}{m_{V_2}}\right)    ,
\end{flalign}
where $a_i$, $b_i$ are arbitrary real numbers ($a_3$, $b_3$ only for massive
vectors). We list below the positivity bounds from each scattering amplitude.
 
\begin{align}
	&ZZ:\nonumber\\
&8 A t_W^4
   \left(F_{S,0}+F_{S,1}+F_{S,2}\right)
   +D t_W^2 \left(-t_W^4 F_{M,3}
   +t_W^2 F_{M,5}-2 F_{M,1}+F_{M,7}\right)
   \nonumber\\
&+(B+C) \left(2 t_W^8 F_{T,9}+4 t_W^4
   F_{T,7}+8 F_{T,2}\right)+8 B
   \left[t_W^4 \left(t_W^4 F_{T,8}+2
   F_{T,5}+2 F_{T,6}\right)+4 F_{T,0}+4
   F_{T,1}\right]\ge0
   \label{boundZZ}
   \\
   \nonumber\\
	&W^\pm W^\pm:\nonumber\\
	&4 A s_W^4 \left(2
   F_{S,0}+F_{S,1}+F_{S,2}\right)
   -8 E s_W^2 F_{M,0}-2
   (E+F) s_W^2 F_{M,1}+F s_W^2
   F_{M,7}
   \nonumber\\&
   +(4 B+6 C) F_{T,2}+16 B F_{T,0}+24 B
   F_{T,1}\ge0
   \\
   \nonumber\\
	&W^\pm W^\mp:\nonumber\\
	&4 A s_W^4 \left(2
   F_{S,0}+F_{S,1}+F_{S,2}\right)
   -2 (G-E) s_W^2 F_{M,1}+8
   E s_W^2 F_{M,0}+G s_W^2 F_{M,7}
   \nonumber\\&
   +(4 B+6 C) F_{T,2}+16 B F_{T,0}+24 B F_{T,1}\ge0
   \\
   \nonumber\\
	&W^\pm Z:\nonumber\\
	&4 A c_W^2 t_W^4
   \left(F_{S,0}+F_{S,2}\right)
   +t_W^2 \left(D-H s_W^2\right)\left(F_{M,7}-2 F_{M,1}\right)
   -H c_W^2 t_W^4
   \left(t_W^2 F_{M,3}+F_{M,5}\right)
   \nonumber\\&
   +4 B \left(t_W^4 F_{T,6}+4 F_{T,1}\right)
   +C \left(t_W^4 F_{T,7}+4  F_{T,2}\right)\ge0
\end{align}
\begin{align}
	&Z\gamma:\nonumber\\
&B \left[
32 c_W^4 \left( F_{T,0}+ F_{T,1}\right)
-16 c_W^2 s_W^2 F_{T,5}
+4(c_W^2-s_W^2)^2F_{T,6}
-F_{T,7} +8 s_W^4 F_{T,8}\right]
   \nonumber\\&
   +(B+C)
   \left[ ( c_W^2 -s_W^2)^2 F_{T,7}+8 c_W^4 F_{T,2}+2 s_W^4 F_{T,9}\right]
   -H c_W^2 s_W^2 \left(2
   F_{M,1}+F_{M,3}+F_{M,5}-F_{M,7}\right)
   \ge0
   \\
   \nonumber\\
	&W^\pm \gamma:\nonumber\\
	& 4 B \left(4 F_{T,1}+F_{T,6}\right)+C \left(4
   F_{T,2}+F_{T,7}\right)-H s_W^2 \left(2 F_{M,1}+F_{M,3}-F_{M,5}-F_{M,7}\right)\ge0
   \\
   \nonumber\\
   &\gamma\gamma:\nonumber\\
   &(B+C) \left(4 F_{T,2}+2 F_{T,7}+F_{T,9}\right)+4 B
   \left(4 F_{T,0}+4 F_{T,1}+2 F_{T,5}+2
   F_{T,6}+F_{T,8}\right)\ge0   , 
   \label{boundgaga}
\end{align}
where
\begin{align}
	s_W\equiv \sin\theta_W,\quad
	c_W\equiv \cos\theta_W,\quad
	t_W\equiv \tan\theta_W,
\end{align}
$\theta_W$ being the weak angle and we have defined
\begin{align}
	\begin{aligned}
&A\equiv a_3^2 b_3^2,
\\
&B\equiv \left(a_1 b_1+a_2 b_2\right){}^2,
\\
&C\equiv \left(a_1^2+a_2^2\right) \left(b_1^2+b_2^2\right),
\\
&D\equiv a_3^2 \left(b_1^2+b_2^2\right)+\left(a_1^2 +a_2^2\right) b_3^2,
\end{aligned}
\qquad
\begin{aligned}
&E\equiv a_3 b_3 \left(a_1 b_1+a_2 b_2\right),
\\
&F\equiv \left(a_1 b_3-a_3 b_1\right){}^2+\left(a_2 b_3-a_3 b_2\right){}^2,
\\
&G\equiv \left(a_3 b_1+a_1 b_3\right){}^2+\left(a_3 b_2+a_2 b_3\right){}^2,
\\
&H\equiv a_3^2 \left(b_1^2+b_2^2\right)	  .
\end{aligned}
\end{align}
The above constraints must hold for arbitrary real values of $a_i$ and $b_i$.
More general positivity bounds can be obtained by considering generic complex polarizations \cite{Bi:2019phv}.

For completeness, we also give the dimension-6 contributions to the positivity inequalities in the Warsaw basis:
\begin{align}
	&WZ:\nonumber\\
	&-a_3^2b_3^2s_W^4c_W^2\left(c_WC_{\varphi D}+4s_WC_{\varphi
	WB}\right)^2
	-36(a_1b_1+a_2b_2)^2e^2s_W^2c_W^2C_{W}^2
	+\mbox{dim-8 terms}
	\ge0
	\\
	&WW:\nonumber\\
	&-a_3^2b_3^2s_W^2c_W^4C_{\varphi D}^2
	-e^2c_W^2 36s_W^2(a_1b_1+a_2b_2)^2 C_W^2
	+\mbox{dim-8 terms}
	\ge0
	\\
	&W\gamma:\nonumber\\
	&-(a_1b_1+a_2b_2)^2C_W^2
	+\mbox{dim-8 terms}
	\ge0
\end{align}
Other channels do not lead to dimension-6 contributions in the results.  As we
can see, the dimension-6 contributions to the left-hand side of the positivity
conditions are negative definite.

\subsection{SM loop contribution}

Here, we compute explicitly the SM loop contribution in the
$\gamma\gamma$ channel as an example, and show that it is negligible
once the low energy discontinuities are subtracted out, as in the r.h.s.~of the
dispersion relation in Eq.~(4).  This is most easily done using Eq.~(5), where
one can see that the remaining contribution of the SM loops comes from the
discontinuities at energies scales higher than $\epi\Lambda$, where the
integrand of the dispersion relation decays as either $1/s^2$ or
$1/s^3$. More explicitly, the one loop SM contribution to
$f_{\epi\Lambda}$ can be computed via the optical theorem using the
tree level total cross section $\gi\gi\to X$:
\begin{flalign}
f^{\rm sm,{\gi\gi}}_{ab,\epi\Lambda}(0) = &  \int^{\infty}_{(\epi\Lambda)^2} \frac{2\ud s}{\pi }   \frac{ {\rm Im}A_{ab}^{\rm sm,\gi\gi}(s)}{ s^3} 
\nonumber\\
=&
\int^{\infty}_{(\epi\Lambda)^2} \frac{2\ud s}{\pi s^3} \sqrt{(s-M_+^2)(s-M_-^2)}
\sum_{X}\sigma_{ab}^{\rm sm}(\gamma\gamma\to X)(s)   ,
\end{flalign}
where we have restricted to the crossing symmetric amplitudes with
$a=(a_1,a_2,a_3)$ and $b=(b_1,b_2,b_3)$ denoting the polarizations. $X$ stands
for possible final states in the SM, which at the tree level includes
$\gamma\gamma\to f\bar f$ (fermion and anti-fermion) and $\gamma\gamma\to
W^+W^-$. To leading order in $(\epi\Lambda)^{-2}$ they are given respectively by
\begin{flalign}
	f^{WW,\gi\gi}_{ab,\epi\Lambda}(0)
	=\frac{16\alpha^2}{(\epi\Lambda)^2m_W^2}
	\left[
		\left(a_1^2+a_2^2\right) \left(b_1^2+b_2^2\right)
	\right]+\mathcal{O}\left[(\epi\Lambda)^{-4}\right]  
\end{flalign}
and 
\begin{flalign}
	f^{ff,\gi\gi}_{ab,\epi\Lambda}(0)
	=&N_cQ^4\frac{2\alpha^2}{(\epi\Lambda)^4}
	\left[
		2\left(a_1^2+a_2^2\right) \left(b_1^2+b_2^2\right)
		\log\frac{(\epi\Lambda)^2}{m_f^2}
		+(a_1^2+a_2^2)(b_1^2+b_2^2)-4(a_1b_1-a_2b_2)^2
	\right]   +\mathcal{O}\left[(\epi\Lambda)^{-6}\right]   .
\end{flalign}
The energy scales that are probed at the LHC for the most constraining high
mass $VV$ pairs are around 1.5-2 TeV \cite{CMS:2018ysc,Sirunyan:2017fvv}, so we
expect the EFT to be valid up to this scale and take $\epi\Lambda=2\,$TeV.  Therefore
the dominant contribution comes from the $\gi\gi\to W^+W^-$ scattering, which
gives $f^{WW,\gi\gi}_{ab,\epi\Lambda}(0)=0.038\,$TeV$^{-4}$ with $|a|$ and $|b|$
normalized to 1.

In comparison, the typical contributions to
$f_{\epi\Lambda}$ from the dim-8 EFT operators are much larger. 
In the convention of \cite{Eboli:2006wa,Eboli:2016kko} which is often
used in experimental analyses, their typical contributions are of order
$f_i/\Lambda^4$.  The current limits on $f_i$ span a few orders of magnitude,
but even the most constraining ones are around $\mathcal{O}(1)
(\Lambda/\mathrm{TeV})^4$.  So the EFT contribution from each operator
is expected to be around $\mathcal{O}(1)$ TeV$^{-4}$, which means the SM
contribution to $f_{ab,\epi\Lambda}^{\gamma\gamma}$ is negligible.  
For example, in the $\gamma\gamma (a\parallel b)$ channel the largest
contribution is from $f_{T,8}$, which gives $9.7\ \mathrm{TeV}^{-4}$,
and the smallest one comes from $f_{T,2}$, which gives $0.10\ \mathrm{TeV}^{-4}$.
All other contributions vary within this range.\\

\subsection{$t$-channel poles}

In the derivation of the improved positivity bounds, the photon $t$-channel
poles, which blow up in the forward limit, are subtracted along with other low
energy poles. Here we explain why this is justified despite of the forward
limit blow-ups.

Consider $B_{\epi\Lambda}(s_p)$ defined in Eq.~(\ref{eq:4}). 
The $WW\to WW$ channels (and only the $WW\to WW$ channels) contain a tree level $t$-channel photon exchange
diagram, which leads to a $t$-channel pole in the forward limit.
However, extended analyticity and other arguments allow one to derive the
positivity bound away from the forward limit for $t=\delta$ and we can take
$\delta$ to be small, as a regulator. (Technically speaking, the non-forward
positivity bounds will generically have additional terms away from definite
transversity scatterings; however, these terms are suppressed by at least
$\delta^{1/2}$ as $\delta\to 0$.) In the SMEFT, the residue of the $t$-channel
pole has up to linear $s$ dependence. Thus the tree level $t$-channel pole will
vanish after twice subtractions, i.e., after taking ${\ud^2
B_{\epi\Lambda}(s_p)}/{\ud s^2}$, and would not affect the positivity
bounds as we take the $\delta\to 0$ limit. 

The remaining $t$-pole contribution may show up in loop diagrams and can also
be subtracted via the cutting rules, again not affecting the positivity
bounds. Consider the optical theorem for the $WW\to WW$ channels
\begin{equation}
{\rm Im} A(s)=[(s-M_-^2)(s-M_+^2)]^{1/2}\sigma_t(s)>0  .
\end{equation}
The r.h.s.~can hit a $t$-channel pole only when $\sigma_t(s)$ contains the
elastic scattering $\sigma_{WW\to WW}(s)$ channels, which receives contribution
from one or more $t$-channel photons.  
If the BSM theory is weakly coupled,
according to the Cutkosky rules, these channels
correspond to the sum of all diagrams with a $WW$ cut on the l.h.s., which we
denote by $A^{WW}(s)$ and $\sigma_{WW}(s)$, that is,
\begin{equation}
{\rm Im} A^{WW}(s)=[(s-M_-^2)(s-M_+^2)]^{1/2}\sigma_{WW}(s) >0 .
\end{equation}
$A^{WW}(s)$ corresponds to at least one-loop diagrams, so away from the forward limit it is a higher-order effect.
 We
want to make sure that $A^{WW}(s)$ does not spoil the analyticity or become
larger than the tree level, as $t\to 0$. To this end, we subtract the $WW\to
WW$ channels from both sides of the optical theorem:
\begin{equation}
	{\rm Im} \left(A(s)-A^{WW}(s)\right)=[(s-M_-^2)(s-M_+^2)]^{1/2}\sigma_{\cancel{WW}}(s)>0 ,
\end{equation}
where $\sigma_{\cancel{WW}}(s)$ is the sum of cross sections of all channels but $WW\to WW$.
Note that $A^{WW}(s)$ also generates additional contributions to the l.h.s.~from
non $WW$ cut in addition to the $WW$ cut, which
need to be removed from the l.h.s., but this does not affect the final result as these
contributions are higher-order effects without a $t$ pole.
We can then define $B_{\epi\Lambda}(s)$ using $A(s)-A^{WW}(s)$ instead of $A(s)$, and arrive at
formally the same positivity bound. With the $t$-channel poles canceled out, the calculation of ${\ud^2
B_{\epi\Lambda}(s_p)}/{\;\ud s^2}$ at the tree level is not affected, because the non-singular parts of $A^{WW}(s)$ are at
least one loop suppressed. Therefore, the $t$-channel poles can all be canceled out and our results are unchanged to leading order. 

In fact, for a weakly coupled UV completion,  effective
operators are matched on to the leading-order BSM amplitude.  This is because
this leading-order BSM amplitude can only involve heavy particle exchanges or
heavy particle loops, and therefore cannot have a $t$-channel pole.  
One can derive a dispersion relation for this leading order BSM
amplitude directly via the cutting rules (see the discussion around Eq.~(\ref{eq:lhcAtr}) or for more details see \cite{Bi:2019phv}). 
For example,
if the leading order BSM amplitude is at the tree level,
given that it satisfies the Froissart bound, we can run the dispersion
relation argument with it and use the cutting rules to infer that the imaginary
part of it is positive. The same argument applies if the leading order is at a
certain loop level. 

If the BSM theory is strongly coupled, one can circumvent the
$t$-channel pole by taking the $g'\to 0$ limit, where $g'$ is the
SM hyper-charge coupling. Since $g'$ is much smaller than the BSM strong coupling, it is expected that the Wilson coefficients, which encode the effects of BSM heavy states, are not significantly altered by this scaling. 

~\\

\end{widetext}

\bibliography{bib}
\bibliographystyle{unsrt}

\end{document}